\documentclass[conference]{IEEEtran}
\IEEEoverridecommandlockouts
\usepackage{cite}
\usepackage[british]{babel}
\usepackage{amsmath,amssymb,amsfonts}
\usepackage{algorithmic}
\usepackage{graphicx}
\usepackage{textcomp}
\usepackage{xcolor}
\def\BibTeX{{\rm B\kern-.05em{\sc i\kern-.025em b}\kern-.08em
    T\kern-.1667em\lower.7ex\hbox{E}\kern-.125emX}}
\begin{document}

\title{Resilient Communication For Avalanche Response in Infrastructure-Limited Environments}

\author{
\IEEEauthorblockN{Joshua Goulton}
\IEEEauthorblockA{
    \textit{School of Computer Science} \\
    \textit{University of Nottingham} \\
    Nottingham, United Kingdom \\
    psyjg13@nottingham.ac.uk
    }
\and
\IEEEauthorblockN{Milena Radenkovic}
\IEEEauthorblockA{
    \textit{School of Computer Science} \\
    \textit{University of Nottingham} \\
    Nottingham, United Kingdom \\
    milena.radenkovic@nottingham.ac.uk}
}

\maketitle

\begin{abstract}
Delay-Tolerant Networks (DTNs) offer a promising paradigm for maintaining communication in infrastructure-limited environments, such as those encountered during natural disasters. This paper investigates the viability of leveraging an existing national transport system—the Swiss rail network—as a data mule backbone for disseminating critical avalanche alerts. Using The Opportunistic Network Environment (ONE) simulator, we model the entire Swiss rail network and conduct a rigorous comparative analysis of two seminal DTN routing protocols: Epidemic and PROPHET. Experiments are performed in two distinct scenarios: alerts originating from dense urban centres and from sparse, remote mountainous regions. Our results demonstrate that the rail network provides robust connectivity for opportunistic communication in both environments thus validating the integration of DTN principles in remote scenarios. 
\end{abstract}

\section{Introduction}
Delay Tolerant Networks (DTNs) have emerged as a promising solution for ensuring reliable communication in scenarios characterised by intermittent connectivity, such as disaster response, remote area communication, and deep-space exploration \cite{spyropoulos2008, khabbaz2012, burleigh2003}. DTNs utilise a store-carry-forward paradigm \cite{lindgren2004}, allowing nodes to retain messages until they encounter another suitable node, thus ensuring message delivery even in highly disrupted network conditions \cite{crowcroft2008, bindra2012}. This feature is particularly crucial in disaster response scenarios, where timely and reliable communication is vital for minimising human casualties and improving rescue operations. \\

Avalanches represent a significant natural disaster scenario, causing numerous fatalities annually despite limited infrastructural damage. In Switzerland alone, over 150 deaths per year are attributable to avalanche-related incidents, underscoring the urgent need for robust communication systems in mountainous terrains \cite{kaya2023, harvey2008, rauch2024}. The survival rate of avalanche victims dramatically decreases with delayed rescue response, highlighting the critical importance of efficient and resilient emergency communication systems. \\

This study investigates the viability of employing opportunistic DTNs leveraging Switzerland’s existing rail network infrastructure to enhance communication reliability in avalanche emergency scenarios. The Swiss rail network offers a unique and advantageous testbed due to its distinct dual connectivity characteristics: densely interconnected urban regions contrasted by sparsely connected mountainous areas. Such dichotomy provides an ideal environment to comprehensively analyse the performance and limitations of DTNs under varied conditions within a single unified infrastructure. \\

We conduct a comparative analysis of two prominent DTN routing protocols - Epidemic and PRoPHET - under realistic simulation scenarios using the ONE simulator \cite{keranen2009}. Epidemic routing relies on widespread message replication, potentially leading to resource exhaustion and congestion, particularly in densely connected areas \cite{vahdat2000}. Conversely, PRoPHET employs probabilistic forwarding based on historical encounter data, aiming to enhance network efficiency by reducing redundant message transmission \cite{lindgren2004, huang2010}.

\section{Related Work}
Delay Tolerant Networks (DTNs) are designed specifically for environments with intermittent connectivity and high latency, leveraging opportunistic communication principles to dynamically establish temporary communication paths \cite{spyropoulos2008, khabbaz2012, crowcroft2008}. These networks utilise node mobility to store, carry, and forward messages \cite{lindgren2004}, thereby bridging the connectivity gaps inherent in challenging environments.

\begin{figure}[htbp]
    \centering
    \includegraphics[width=\columnwidth]{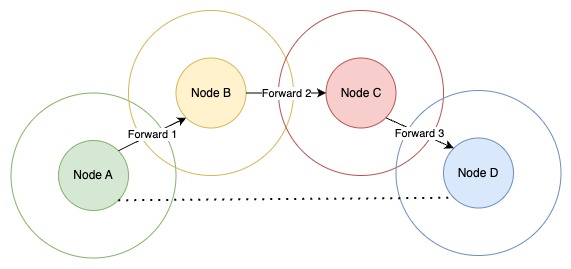}
    \caption{“Store carry forward” connection A-D shown using transitive
                communication\cite{lindgren2004}. (Node shaded area signifies node
                storage, Node ring signifies communication radius)}
    \label{fig:store_carry_forward}
\end{figure}

\subsection{Opportunistic Networking Principles}
Opportunistic networking involves nodes exchanging data opportunistically when in proximity, without the requirement of a continuous end-to-end path \cite{ali2010, lindgren2004}. Unlike traditional networks, DTNs exploit transient contacts and mobility patterns, facilitating effective data dissemination even under disrupted conditions \cite{bindra2012, vahdat2000, huang2010}.

\subsection{Existing Applications of DTNs in Disaster Scenarios}
The adaptability of DTNs has resulted in their application across various disaster scenarios, demonstrating their importance beyond theoretical contexts. Airborne search and rescue missions using unmanned aerial vehicles (UAVs) have successfully utilised DTNs for sharing critical real-time information among mobile rescue teams, enabling efficient operation management in affected regions \cite{bujari2018}.\\

In the context of wildfires, Huynh and Radenkovic developed an Energy-aware Emergency Communication Framework (E3F), which addresses the demands of large-scale disaster environments by intelligently managing energy resources, storage, and forwarding strategies. This approach significantly enhances network reliability and resilience amidst severe disruptions \cite{huynh2017}. Further studies by Huynh and Radenkovic include the implementation of energy-aware opportunistic charging and energy distribution strategies in vehicular edge and fog networks. These strategies optimise energy usage, supporting sustainable operations especially critical when conventional energy infrastructures are disrupted by disasters, further aiding remote scenarios \cite{radenkovic2020energy}.\\

Moreover, heterogeneous networks comprising of drones and vehicles for real-time communication in dynamic disaster scenarios have been utilised, emphasising adaptive networking solutions to maintain persistent communication in highly mobile environments. Such systems ensure critical data dissemination during emergencies \cite{radenkovic2019drones}. Additionally, personal mobile cloud solutions have been proposed and provide a low-cost peer-to-peer architecture that securely facilitates the sharing of sensitive personal data such as health information in disaster scenarios, ensuring both privacy and rapid information dissemination \cite{radenkovic2015mobile}. The MODiToNeS testbed was also developed aiding research groups and providing practical insights into the real-world application of DTNs. This low-cost platform allows researchers and practitioners to prototype and evaluate opportunistic network algorithms in realistic scenarios, bridging the gap between theoretical frameworks and practical implementations \cite{radenkovic2019drones, radenkovic2017fault}.\\

Collectively, these applications demonstrate that DTNs significantly enhance the capabilities and effectiveness of disaster response operations, making them indispensable for future emergency management strategies. This study builds upon such foundational work by applying DTN principles to avalanche response scenarios using the Swiss rail network, thereby contributing novel insights into DTN performance across geographically diverse conditions.

\section{Simulation Design}
To rigorously evaluate the performance of DTN protocols for avalanche response, we designed a set of realistic simulation experiments. This section details the simulation environment, the modelling of the network scenario based on the Swiss rail network, the configuration of nodes and communication interfaces, and the specific experimental tests conducted.\\

\subsection{Simulation Environment}
The experiments were conducted using The Opportunistic Network Environment (ONE) simulator \cite{keranen2009}, a tool specifically designed for evaluating DTN routing and application protocols. The ONE simulator was chosen for its comprehensive features, including its ability to model node movement along predefined paths based on real-world map data, its support for various established DTN routing protocols, and its detailed reporting capabilities for performance analysis. Its Java-based architecture ensures cross-platform compatibility and facilitates the creation of reproducible experimental setups.\\

The geographical and network data for the Swiss rail system were processed and abstracted using the open-source Quantum GIS (QGIS) software  \cite{moyroud2018} with the GRASS GIS plugin \cite{lacaze2018}. This process allowed for the conversion of the complex, real-world rail network topology into a map format compatible with the simulator. The resulting map data were formatted as a Well-Known Text (WKT) file, which the ONE simulator uses to constrain the mobility of train nodes to the actual railway tracks, thereby ensuring a high-fidelity representation of the physical environment.\\

\begin{figure}[htbp]
    \centering
    \includegraphics[width=0.5\textwidth, height=0.25\textwidth]{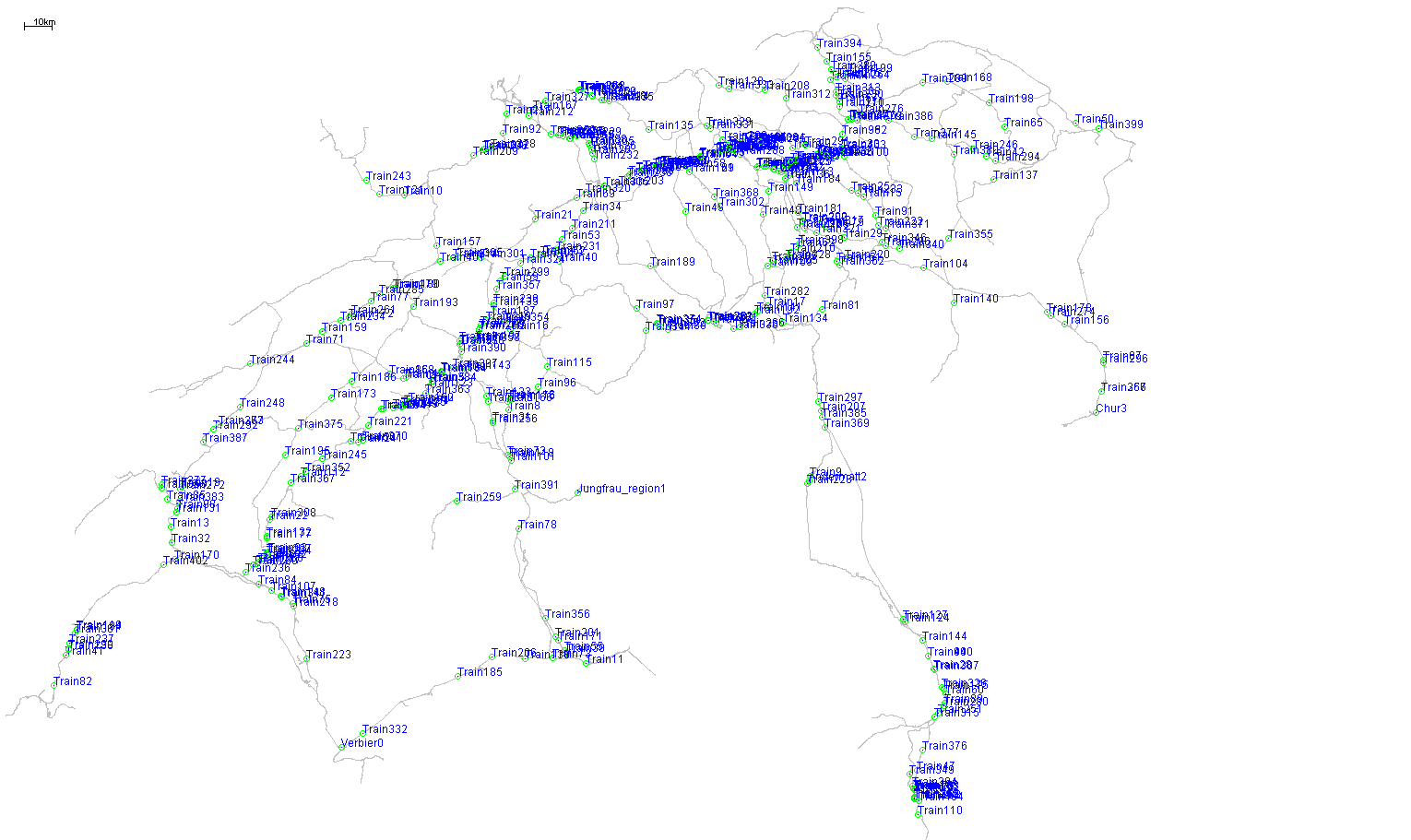}
    \caption{Scenario represented in the One simulator}
    \label{fig:one_map}
\end{figure}

\subsection{Disaster Modelling}

The core of our study is an avalanche response scenario set in Switzerland. This location was deliberately chosen due to its unique geographical dichotomy: densely interconnected urban regions in the Swiss Plateau contrasted with sparsely connected, mountainous regions in the Alps. This contrast provides an ideal testbed to analyse and compare protocol performance under fundamentally different network conditions—high connectivity and frequent contact opportunities versus intermittent connectivity and sparse contacts—all within a single, cohesive infrastructure. The objective is to determine the viability of leveraging the existing national rail network as a data mule for relaying critical emergency alerts from a disaster zone to a central response hub.\\

The simulation environment consists of two primary types of scenarios:

Urban Alert Scenario: An alert is generated from a stationary node located in a major city (Lausanne, Bern, Luzern, or St. Gallen). These locations represent areas with high train traffic and network density.\\

\begin{figure}[htbp]
    \centering
    \includegraphics[width=\columnwidth]{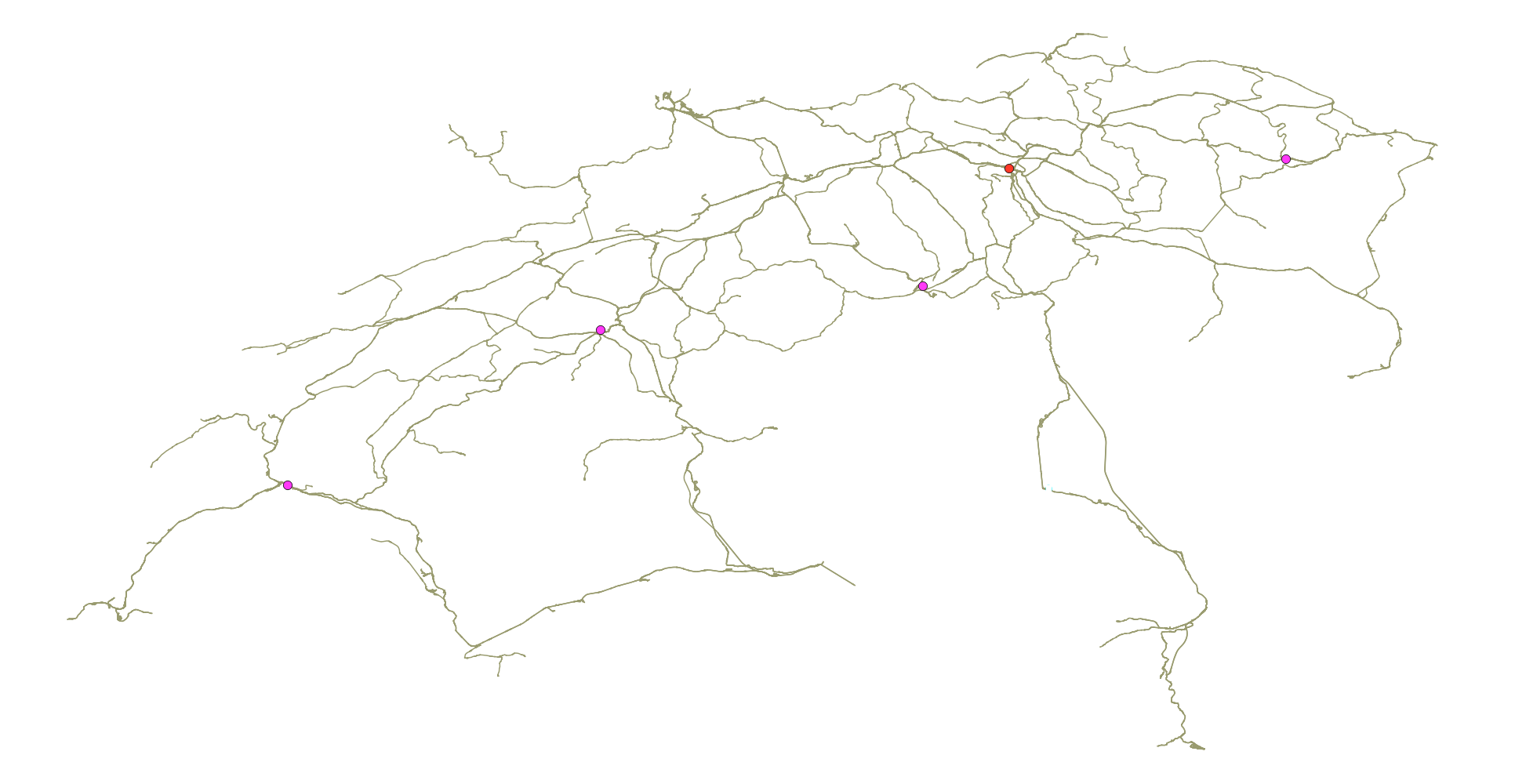}
    \caption{Mapping of city alert sources (Pink) and alert destinations (Red) on the
            Swiss rail map visualised using QGIS. Source destinations from west to east:
            Lausanne, Bern, Luzern, St. Gallen}
    \label{fig:inner_map}
\end{figure}

Mountainous Disaster Scenario: An alert originates from a stationary node in a remote, mountainous region prone to avalanches (Verbier, the Jungfrau region, Andermatt, or Chur). These locations are characterised by lower train traffic and represent connectivity-challenged environments.\\

For all scenarios, the designated destination for the alert message was a fixed node located in Zurich, representing a national emergency response coordination centre.\\

\begin{figure}[htbp]
    \centering
    \includegraphics[width=\columnwidth]{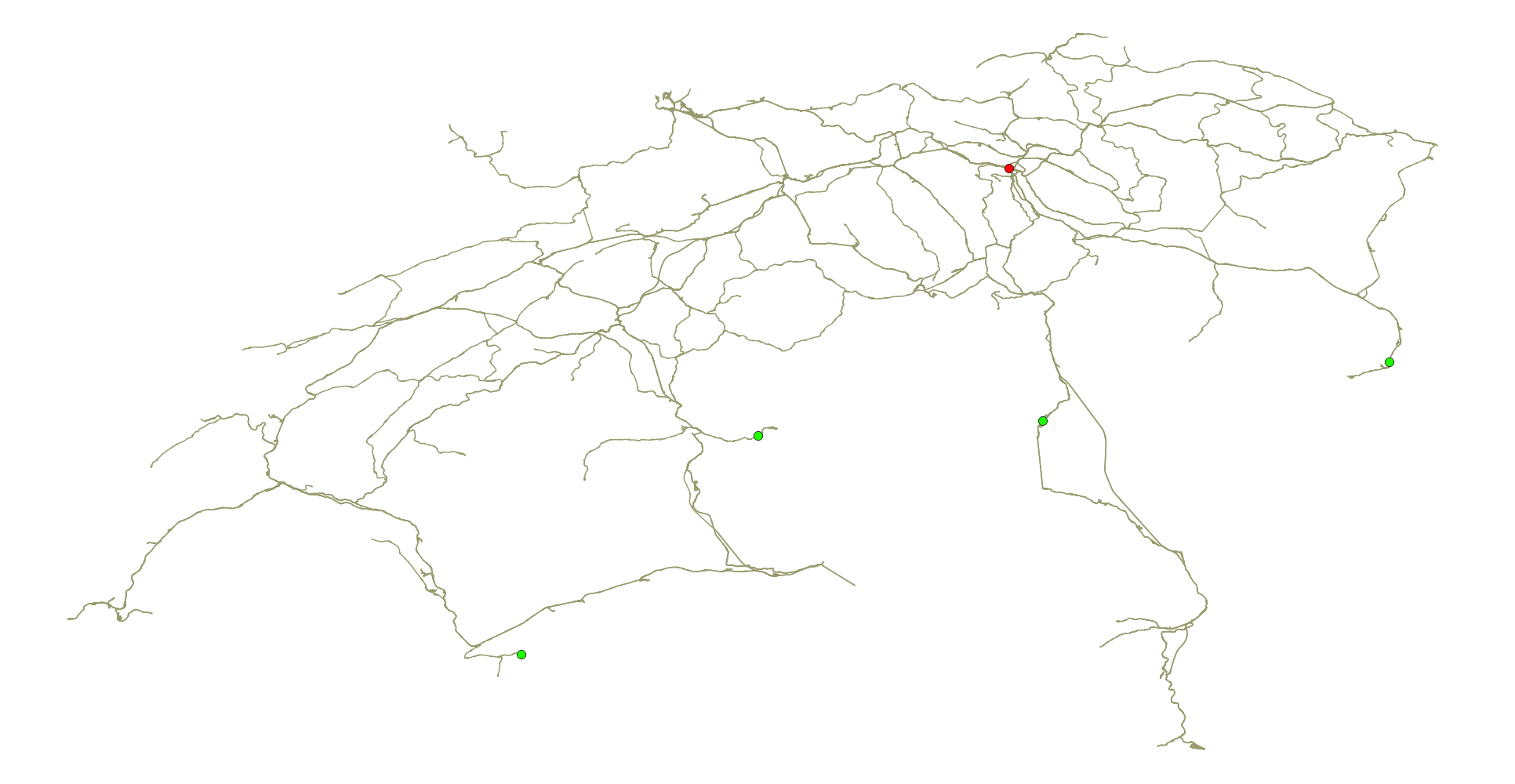}
    \caption{Mapping of mountainous alert sources (Green) and alert destinations
            (Red) on the Swiss rail map visualised using QGIS. Source destinations from west
            to east: Verbier, Jungfrau region, Andermatt, Chur}
    \label{fig:mountain_map}
\end{figure}

\subsection{Node Mobility}
The simulation network is composed of a heterogeneous set of nodes:

Mobile Nodes (Trains): A total of 400 mobile nodes representing trains were configured. These nodes act as the primary data mules, responsible for carrying and forwarding messages across the network. Their movement is governed by the MapBasedMovement model in the ONE simulator, restricted to the paths defined by the Swiss rail network WKT map. The trains follow their predefined routes and schedules, reflecting a realistic operational model where message delivery is purely opportunistic and does not involve altering train paths for data ferrying.

Stationary Nodes (Sources \& Destination): In each experimental run, a single stationary source node was responsible for generating an alert message. An additional stationary node in Zurich acted as the message sink (destination). These nodes do not move and rely entirely on transitive communication with the mobile train nodes to send and receive messages. The placement of these source nodes was varied across the urban and mountainous locations described previously to ensure a comprehensive analysis across different geographical and network conditions.\\

\subsection{Communication Interfaces}
To accurately model real-world communication possibilities, nodes were equipped with two distinct wireless interfaces, employed based on the operational context and proximity:\\

Bluetooth: This interface was configured for short-range, high-speed data exchange (simulating Bluetooth v4/v5 capabilities \cite{bisdikian2001}). It represents the primary communication method between train nodes when they are in close proximity, such as when passing each other on adjacent tracks or waiting at a station. This is particularly relevant in dense urban environments where inter-node contacts are frequent.

LoRa (Long Range): This interface was modelled to reflect the characteristics of LoRaWAN technology \cite{devalal2018}, which provides long-range, low-power communication. It was designated as the communication link between the static mountainous disaster nodes and the mobile train nodes. The rationale is that a fixed sensor node in a remote valley would not be directly on a railway line but could realistically transmit an alert over a distance of several kilometers to a passing train. This dual-interface approach adds a layer of realism, acknowledging that different technologies would be suited for different parts of the disaster response communication chain.\\

\subsection{Testing Parameters}
Each simulation experiment was run for a total of 24 simulated hours (86,400 seconds) to capture a full day's worth of train movement and contact opportunities, allowing for sufficient data collection on message propagation dynamics. A single message, representing a critical avalanche alert, was generated by the source node at the beginning of each simulation. The performance of the routing protocols was evaluated based on their ability to deliver this message to the destination node in Zurich.\\

Two contrasting DTN routing protocols were implemented and compared in each scenario:\\
Epidemic Routing \cite{vahdat2000}: A flooding-based protocol where nodes exchange messages with every peer they encounter, aiming to maximize delivery probability at the cost of high resource overhead.

PROPHET (Probabilistic Routing Protocol using History of Encounters and Transitivity) \cite{lindgren2004}: A probabilistic protocol that uses historical encounter data to calculate delivery predictabilities, forwarding messages only to nodes that have a higher probability of reaching the destination. This is intended to reduce overhead compared to Epidemic routing.\\

By running each of the urban and mountainous scenarios with both protocols, we generated a comprehensive dataset to compare their performance in terms of delivery probability, latency, hop count, and network overhead.\\

\section{Experimental Results}
\subsection{Network Connectivity Analysis}
To establish a baseline for protocol performance, we first analysed the connectivity characteristics of the simulated Swiss rail network. The inherent mobility patterns of the 400 train nodes create a dynamic topology whose density varies significantly between urban and mountainous regions.\\

\begin{figure}[htbp]
    \centering
    \includegraphics[width=\columnwidth]{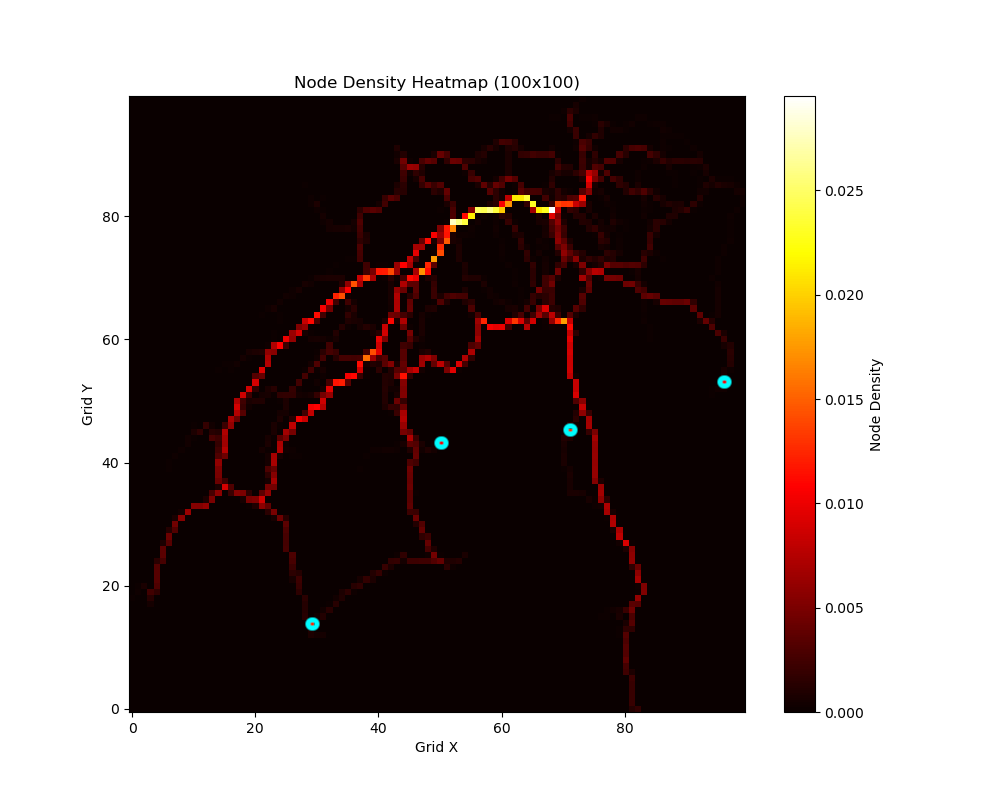}
    \caption{Node density heat map throughout the simulation (blue representing mountain nodes)}
    \label{fig:heatmap}
\end{figure}

A node density heatmap, illustrated in Figure \ref{fig:heatmap}, reveals the spatial distribution of nodes over the 24-hour simulation period. As expected, the primary rail corridors connecting major cities exhibit the highest node density, indicating frequent contact opportunities. Conversely, rail lines extending into the Alpine regions are sparser. However, an interesting phenomenon was observed on certain single-track mountain routes, such as the line passing Andermatt (third blue centre reading left to right). On these constrained paths, nodes spend a significant portion of their travel time, leading to localized areas of high density. This suggests that even in remote regions, if an alert message is successfully passed to a train on such a route, it can be rapidly disseminated among other nodes travelling along the same corridor.\\

\begin{figure}
    \centering
    \includegraphics[width=\columnwidth]{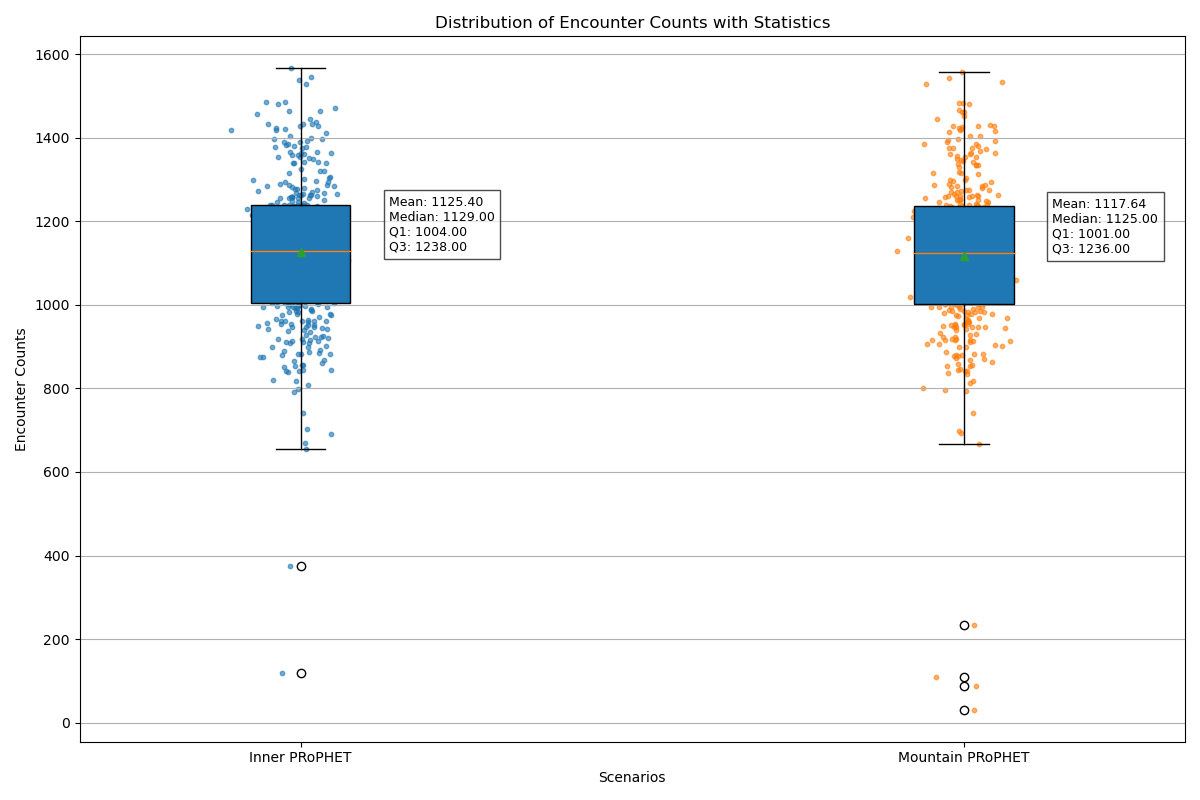}
    \caption{Total distribution of encounters in the two environments}
    \label{fig:Node_encounters}
\end{figure}

This observation is further substantiated by the node encounter statistics in Figure \ref{fig:Node_encounters}. Despite the geographical dispersion, the difference in average node encounters between the two scenarios was minimal. In the Urban scenario, nodes had a mean of approximately 1125 encounters over 24 hours. In the Mountainous scenario, the mean was only slightly lower, at approximately 1117 encounters, a difference of less than 1\%. This indicates that while the network is sparser in mountainous regions, the connectivity is still remarkably high, providing a robust foundation for opportunistic data dissemination.\\

\subsection{Comparative Protocol Performance}

We evaluated the performance of the Epidemic and PROPHET protocols across both Urban and Mountainous scenarios. The key performance indicators are summarized in Table I.\\

\begin{table}[!t]
\caption{Summary of Protocol Performance Metrics}
\label{tab:performance_metrics}
\centering
\resizebox{\columnwidth}{!}{%
\begin{tabular}{|l|l|c|c|c|}
\hline
\textbf{Scenario} & \textbf{Protocol} & \textbf{Deliv. Prob.} & \textbf{Avg. Hops} & \textbf{Latency} \\
\hline
Urban       & Epidemic & 20.53\% & 19.43 & 0.69 \\
            & PROPHET  & 25.33\% & 11.49 & 0.68 \\
\hline
Mountainous & Epidemic & 22.45\% & 12.91 & 0.90 \\
            & PROPHET  & 29.24\% & 11.31 & 1.00 \\
\hline
\end{tabular}
}
\end{table}

Delivery Probability: This metric represents the fraction of messages successfully delivered to the destination based on intermittent signals from the source. PROPHET consistently outperformed Epidemic routing in both scenarios. In the Urban scenario, PROPHET achieved a delivery probability of 25.33\%, compared to 20.53\% for Epidemic. The margin was even wider in the Mountainous scenario, where PROPHET achieved 29.24\% delivery probability versus 22.45\% for Epidemic. This superior performance is attributable to PROPHET's intelligent forwarding mechanism. By prioritizing nodes with a higher historical probability of reaching the destination, it mitigates network congestion and reduces the number of dropped packets that result from buffer overflows, a common issue with Epidemic's indiscriminate flooding approach.\\

Average Hop Count: This metric measures the average number of intermediary nodes a message traverses from source to destination. PROPHET demonstrated significantly greater efficiency, requiring substantially fewer hops. In the Urban scenario, PROPHET messages averaged 11.49 hops, while Epidemic messages required 19.43 hops. A similar trend was observed in the Mountainous scenario (11.31 for PROPHET vs. 12.91 for Epidemic). The high hop count for Epidemic, especially in the dense urban environment, is a direct consequence of its flooding strategy, which causes messages to propagate through numerous redundant relays. PROPHET's selective forwarding, in contrast, facilitates more direct and efficient routing paths.\\

Scaled Latency: To provide a fair comparison of delivery times across scenarios with different geographical distances, latency was normalized against the maximum observed delivery time in any experiment. A value of 1.00 represents the highest latency recorded (Mountainous PROPHET). The results show that within each scenario, the latency difference between protocols is negligible. However, the latency is predictably higher for the Mountainous scenarios due to the greater average physical distance from the source nodes to Zurich. The key finding here is that the use of DTNs on the existing rail network does not introduce a prohibitive latency penalty for remote regions when compared to the increased distance hence, the system remains effective.\\

\subsection{Network Overhead and Resource Utilization}
To assess the resource efficiency of each protocol, we analysed the total buffer occupancy across all network nodes over time. This metric is critical, as excessive buffer usage leads to packet loss and strains the resources of network nodes.\\

\begin{figure}
    \centering
    \includegraphics[width=\columnwidth]{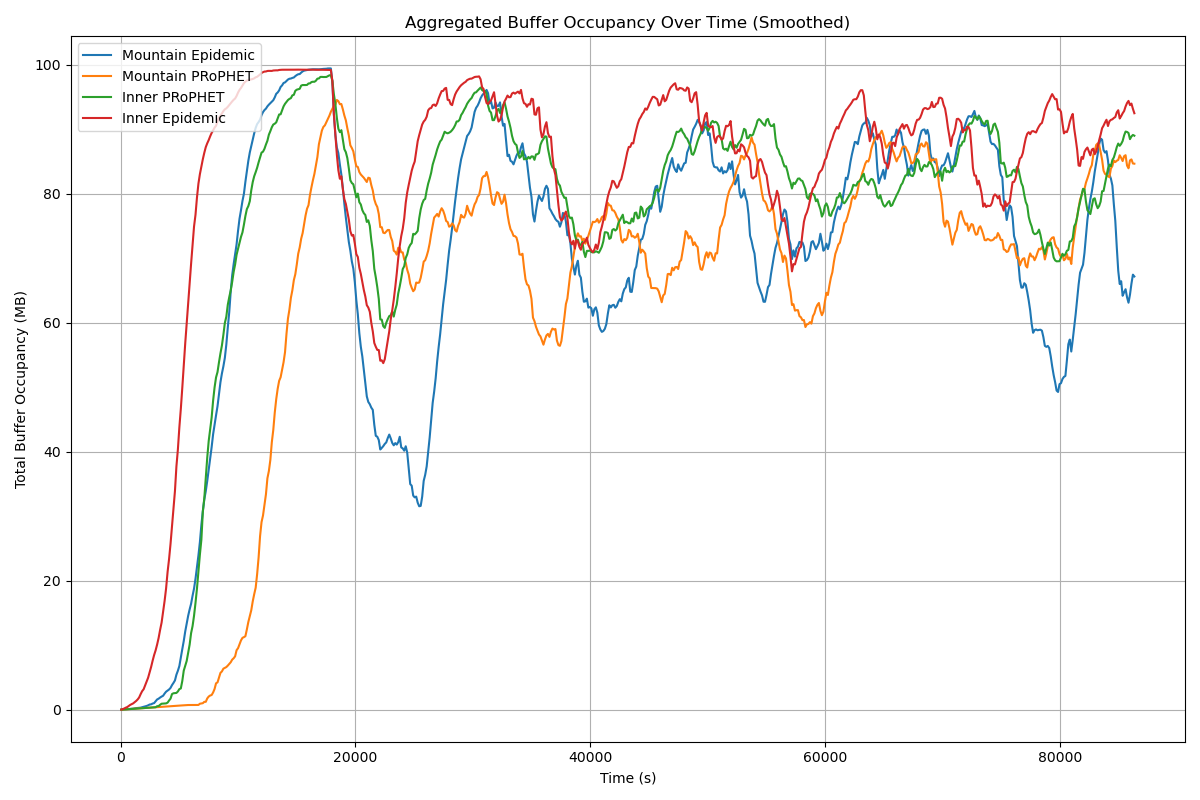}
    \caption{Buffer occupancy throughout the 24 hour simulation}
    \label{fig:buffer}
\end{figure}

The aggregated buffer occupancy, depicted over the 24-hour simulation period in Figure \ref{fig:buffer}, shows distinct patterns for each protocol and scenario. Epidemic routing consistently resulted in higher buffer utilization due to its message replication strategy. In the dense Urban scenario, Epidemic's buffer occupancy was the highest observed, averaging 82.82 MB. The frequent inter-node contacts lead to relentless message copying, keeping buffers near saturation. The variance was also the lowest (2.79 MB), indicating that the high buffer load was uniformly distributed across all nodes, leaving little capacity margin system-wide.\\

In contrast, PROPHET demonstrated more moderate buffer usage. In the Urban scenario, it averaged 75.77 MB, effectively reducing the network load compared to Epidemic. In the Mountainous scenario, the difference was also clear, with Epidemic averaging 69.73 MB and PROPHET averaging 64.69 MB.\\

An important observation is the variance in buffer occupancy. PROPHET consistently exhibited a higher variance (e.g., 7.21 MB in the mountains vs. Epidemic's 3.2 MB). This indicates that its probabilistic forwarding creates a more uneven load distribution. Nodes identified as having high delivery predictability become "hubs" and carry a larger portion of the network traffic, while less "useful" nodes on a less central train route carry fewer messages. While this leads to overall network efficiency, it highlights a potential risk of creating localized congestion points on critical nodes. Nonetheless, across all analysed metrics, PROPHET's strategy proves to be superior, delivering more messages with fewer hops and lower overall resource consumption.\\

\section{Future Work}
Our findings indicate that even a basic level of historical context, as used by PROPHET, significantly enhances performance. The next logical step is to evaluate more advanced routing protocols that leverage richer forms of social or contextual information.\\

Socially-Aware Protocols: Protocols such as MaxProp \cite{burgess2006} and the Socially-Aware Adaptive DTN (SAAD) routing protocol \cite{ullah2022} utilize more complex social metrics, like centrality and community structures, to make forwarding decisions. Given the structured and repetitive nature of a rail network, these protocols could identify critical hub stations or highly trafficked train routes, leading to further optimization of message delivery paths and resource management. Mobility-Aware Protocols: The predictable trajectories of trains make mobility-aware protocols particularly suitable. For instance, DirMove \cite{gupta2016} uses the direction of node movement to inform routing decisions. In our scenario, this could prevent messages from being forwarded to a train moving away from the destination, a simple yet highly effective optimization.\\

The unique and deterministic nature of a national rail network offers an opportunity to design a novel protocol that goes beyond historical observation. We propose the development of a Schedule-Aware Routing (SAR) protocol. In this paradigm, train nodes would not need to infer connectivity patterns over time; instead, they could be updated with real-time or short-term schedule information at stations or through a separate control channel. This a priori knowledge of future encounters would allow for near-optimal forwarding decisions, drastically reducing the overhead associated with discovery and probabilistic calculations. Such a protocol could represent the pinnacle of efficiency in this type of structured DTN.\\

While this study focused on delivery performance, real-world deployments of static sensor nodes are critically constrained by energy. Future work should incorporate energy consumption into the simulation models. Evaluating protocols within an energy-aware framework, such as the E3F \cite{huynh2017}, would be essential for designing a sustainable system. This includes exploring mechanisms for opportunistic charging or low-power communication strategies to maximize the operational lifespan of disaster-monitoring nodes travelling within the network on the same infrastructure but crucially separate from the trains themselves.

\bibliographystyle{IEEEtran}

\vspace{12pt}

\end{document}